\def\beq{\begin{equation}}
\def\eeq{\end{equation}}
\def\bea{\begin{eqnarray}}
\def\eea{\end{eqnarray}}

\def\th{_{_{\rm th}}}
\documentstyle[prl,aps]{revtex}
\begin{document}
\title{Wormholes in spacetime with torsion}
\author{Luis A. Anchordoqui \thanks{Electronic address: 
doqui@venus.fisica.unlp.edu.ar}}

\address{Departamento de F\'{\i}sica, Universidad Nacional de La Plata\\
C.C. 67, 1900, La Plata, Buenos Aires, Argentina}
\maketitle

\begin{abstract}
Analytical wormhole solutions in $U_4$ theory are presented. 
It is discussed whether the extremely short range repulsive
forces, related to the spin angular momentum of matter, could be the 
``carrier'' of the exoticity that threads the wormhole 
throat.
\end{abstract}




\section{Introduction}

Wormhole physics has crept back into the literature since the
analysis of classical traversable wormholes performed by Morris and
Thorne \cite{motho}. 
Intuitively speaking such wormholes are tunnels linking widely
separated regions of space-time from where in-going causal curves can pass 
through and become out-going on the other side.
The most striking feature of this unconventional arena is the possibility of 
constructing time machines. It has been shown that generic relative
motions of the two wormhole's mouths, or equivalently generic gravitational
redshifts at the mouths due to external gravitational fields might 
produced closed timelike curves \cite{mothoyu}. 
A fundamental
difficulty associated with traversable wormholes, even in the
absence of closed timelike curves, is the requirement of 
gravitational sources that 
violate the weak energy condition (WEC) in order to support them. 
Geometrically this could be understood as a breakdown of 
Hawking-Penrose singularity theorems \cite{hawpen}. Instead of becoming 
singular the timelike or null geodesic congruences diverge at the 
wormhole throat. Attempts to get around the violation of WEC for 
static wormholes in
General Relativity (GR)
have led some to look at wormholes in non-standard gravity theories, such
as $R + R^2$ theories \cite{hoch}, 
Moffat's non-symmetric theory
\cite{mof}, Einstein-Gauss-Bonnet theory \cite{bha}, and Brans-Dicke
(BD) theory
\cite{bdwh}. Nevertheless, for static wormholes, the violation of WEC is a
necessary consequence, no matter how many extra degrees of freedom
one wishes to endow upon the theory. Thus, in the case of static
BD wormholes,
the BD scalar ends up as the ``carrier'' of
exoticity. 
The bottom line is that WEC is violated
for all static wormholes \cite{vishoch}. Within dynamical wormholes,
it was shown that multiplying  a static spherically symmetric line element by 
an overall time dependent conformal factor it is possible (depending on 
the explicit form chosen for the conformal factor) to either postpone the 
violation of WEC, to relegate it to the past, or else to restrict its 
violation to short intervals of time \cite{kar,ewh}. 
The bizarre gravitational interactions of Einstein-Cartan (EC) theory have 
proved to be usefull in the prevention of singularities \cite{Trau}. Thus, 
these interactions emerge as the most attractive mechanism to obtain 
the ``flash'' 
of ``exotic'' matter required to thread an evolving wormhole geometry.
The present letter reports our findings in regard to spherically 
symmetric classical wormhole solutions in $U_4$ theory.

\section{Evolving wormholes endowed with torsion}

The differentiable 
spacetime manifold of EC theory  has a non-symmetric
affine conection\footnote{In what follows, 
greek indices $\mu,\nu,\dots,$ 
run from 0 to 3; square brackets 
denote antisymmetrization while parenthesis symetrization, 
and we set $G$ = $c$ = 1.}. 
Its antisymetric part, the torsion tensor 
$S_{\mu\nu}\,^\eta \equiv \Gamma^\eta_{[\mu\nu]}$,  is
linked to the spin angular momentum of matter $\tau_{\mu\nu}\,^\eta$. The field
equations of $U_4$  theory are found to be \cite{hel},
\beq
R_{\mu\nu} - \frac{1}{2} g_{\mu\nu} R_\eta\,^\eta = 8 \pi \Sigma_{\mu\nu}
\label{!}
\eeq
\beq
S_{\mu\nu}\,^\eta + \delta_\mu^\eta S_{\nu\alpha}\,^\alpha - \delta_\nu^\eta S_{\mu\alpha}\,^\alpha = 8 \pi
\tau_{\mu\nu}\,^\eta
\label{!!}
\eeq
where 
$\delta_\mu^\nu$ is the Kronecker delta, and $g_{\mu\nu}$ the metric
tensor with signature -2. $R_{\mu\nu}=R_{\eta\mu\nu}\,^\eta$, 
with $R_{\mu\nu\eta}\,^\alpha$ the
curvature tensor of the Riemann-Cartan connection $\Gamma_{\mu\nu}^\eta
=\{_{\mu\nu}^\eta\}+
S_{\mu\nu}\,^\eta-S_\nu\,^\eta\,_\mu+S^\eta\,_{\mu\nu}$, being 
$\{_{\mu\nu}^\eta\}\,$ 
the Christoffel symbol of the metric. $\Sigma_{\mu\nu}$ is the 
stress-energy tensor,
\beq
\Sigma_{\mu\nu} = (\rho + \Pi) u_\mu u_\nu - \Pi g_{\mu\nu} + \Delta
\Pi \left[ c_\mu c_\nu + \frac{1}{3} (g_{\mu \nu} - u_{\mu} u_\nu) \right]
+ 2 \,q \,c_{(\mu}u_{\nu)}
\eeq
with $\rho$ the energy density, $\Pi =  1/3\, (p_r + 2 p_\perp)$, 
the isotropic pressure, and $q$ the energy flux. 
The anisotropic pressure, $\Delta \Pi$, is the
difference between the local radial and lateral stresses $p_r$,
$p_{\perp}$. $u^\mu$ is the four velocity and $c^{\mu}$ is a space-like 
unit vector in the radial direction.
If one substitutes Eq. (\ref{!!}) in Eq.
(\ref{!}), after a bit of algebra one arrives at the combined field equation,
\beq
R^{\mu\nu}(\{\}) - \frac{1}{2} g^{\mu\nu} R_\eta\,^\eta (\{\}) = 8 \pi 
\tilde{\sigma}^{\mu\nu}
\label{E}
\eeq
with
\beq
\tilde{\sigma}^{\mu\nu} \equiv \sigma^{\mu\nu} + 8 \pi \,
[-4\tau^{\mu\eta}\,_{[\alpha}\tau^{\nu\alpha}\,_{\eta]} - 2 
\tau^{\mu\eta\alpha} \tau^\nu\,_{\eta\alpha} +
\tau^{\eta\alpha\,\mu} \tau_{\eta\alpha}\,^\nu + 1/2\, g^{\mu\nu} \,
(\,4 \,\tau_\xi\,^\eta\,_{[\alpha}
\tau^{\xi\alpha}\,_{\eta]} + \tau^{\xi\,\eta\alpha} \tau_{\xi\,
\eta\alpha}\,)]
\label{S}
\eeq
$\{\}$ means that the quantities have been computed for the
Riemannian part, $\{^\eta_{\mu\nu}\}$, of the affine connection and are the
same as in GR. $\sigma^{\mu\nu}$ is the metric energy
momentum tensor, 
\beq
\sigma^{\mu\nu} = \Sigma^{\mu\nu} - \nabla^{\star}_\lambda
(\tau^{\mu\nu\lambda} - \tau^{\nu\lambda\mu} + \tau^{\lambda \mu \nu})
\eeq
with $\nabla^{\star}_\lambda = \nabla_\lambda \,+ 2\,
S_{\lambda\rho}\,^\rho$, being $\,\nabla_\lambda$ the covariant
derivative with respect to the affine connection $\Gamma_{\lambda\rho}^\mu$.
The spin tensor is given by, 
\beq
\tau_{\mu\nu}\,^\eta = s_{\mu\nu} u^\eta
\eeq
with the constraint,
\beq
s_{\mu\nu} u^\nu = 0  
\eeq
where $s_{\mu\nu}$ is the spin density. The quantities 
$\sigma^{\mu\nu}$ and $\tau_{\mu\nu}\,^\lambda$ are microscopically 
fluctuating, thus, one has to compute an average of Eq. (\ref{S}) in 
order to obtain an equation for bulk matter.
In taking the average of a spherically symmetric, isotropic system 
of randomly oriented spins, it is found that the spin average itself $<
s_{\mu\nu} >
= 0 $, while the spin square terms $<s _{\mu\nu}s^{\mu\nu} > \neq 0$. 
Hereafter we shall drop the averaging signs $<>$, but bear in mind that we 
are dealing with microscopically averaged quantities.

Let the spacetime metric to assume the diagonal form, 
\beq
ds^2 = \Omega (t) \, \{ \, e^{2\,\Phi(r)^{^{\pm}}}\,\, dt^2 - e^{2\,
\Lambda(r)^{^{\pm}}} \,\, dr^2 - r^2\,\,
d\Omega_2^2 \,\}
\eeq
with $\Omega (t)$ a conformal factor finite and positive defined
throughout the domain of $t$, $\Phi$ the redshift function, $\Lambda$ the 
shape-like 
function, and $d\Omega_2^2$, 
the $S^2$ line element. The signs $+$, $-$, are used to represent quantities 
in the upper and lower universes respectively.

In the spirit of \cite{karsah}, we make the Ansatz $\Phi = -
\alpha/r$, where $\alpha$ is a positive constant to be determined. 
This choice guarantees that the redshift function $\Phi$ is finite 
everywhere, and consequently there are no event horizons.
The application of the
field equations (\ref{E}) leads to the following expressions,  
\begin{equation}
 \frac{3 \,\dot{\Omega}^2 \,e^{2\alpha/r}}{4\,\Omega^3} + \frac{2 \,\Lambda' \,
e^{-2\Lambda}}{r \, \Omega} + \frac{1}{r^2 \, \Omega} - \frac{e^{-2 \Lambda
}}{r^2 \,\Omega} = 8 \pi \rho - 16 \pi^2 s^2 = 8 \pi \tilde{\rho}
\label{e1}
\end{equation}
\begin{equation}
\frac{3\,\dot{\Omega}^2 \,e^{2\alpha/r}}{4\,\Omega^3} + \frac{ 2 \,\alpha
\, e^{-2\Lambda}}{r^3\, \Omega} - \frac{\ddot{\Omega} \, e^{2\alpha/r}}
{\Omega^2}
- \frac{ 1}{\Omega \, r^2} + \frac{ e^{-2\Lambda} }{\Omega \, r^2} = 
8 \pi p_r - 16 \pi^2 s^2 = 8 \pi \tilde{p_r}
\label{e2}
\end{equation}
\begin{equation}
 \frac{3 \, \dot{\Omega}^2 \, e^{2\alpha/r}}{4\, \Omega^3} 
- \frac{\ddot{\Omega}\, e^{2\alpha/r}}{\Omega^2}
- \frac{e^{-2\Lambda} \,\alpha}{r^3\, \Omega} - \frac{\Lambda'\,e^{-2 \Lambda}}
{r \, \Omega} + \frac{\alpha^2 \,e^{-2 \Lambda}}{ r^4 \,\Omega} - 
\frac{ \alpha\, \Lambda'\,
e^{-2 \Lambda}}{\Omega \, r^2} =  8 \pi p_{\perp} - 16 \pi^2 s^2 = 8
\pi \tilde{p_{\perp}}
\label{e3}
\end{equation}
\begin{equation}
\frac{ \dot{\Omega} \,\alpha }{r^2
\, \Omega^2}
= 8 \pi q \,\, e^{\Lambda} \, e^{ \alpha/r}
\label{flujo}
\end{equation}
with $s^2 = 2 s_{\mu \nu} s^{\mu \nu}$ the square of the spin density.
In the expressions given above, dashes denote derivatives with respect
to $r$ while dots, derivatives 
with respect to $t$. 
Following the procedure presented in \cite{ewh}, we shall look for
analytical wormhole solutions independet of the conformal factor
$\Omega(t)$. In order to do so, let us impose the constraint,
\begin{equation}  
\rho -  \Pi = \frac{ \Upsilon (t) \, e^{2\alpha/r}}{ \Omega(t)} 
\end{equation}
or equivalentely, 
\begin{equation}
 \frac{\ddot{\Omega}\,e^{2\alpha/r } }{\Omega^2} +
\frac{e^{-2\Lambda}}{3\, \Omega} \left[ \Lambda' \left(
\frac{4}{r} + \frac{\alpha}{r^2} \right)  +  \frac{2e^{2\Lambda}}{r^2} -
\frac{2}{r^2} - \frac{\alpha^2}{r^4} \right]  
 =   \Upsilon (t)  \frac{e^{2\alpha/r}}{\Omega(t)}.
\label{traza}
\end{equation}
Actually, the definition of $\Upsilon$ is needed
so as to obtain the complete behavior of the equation 
of state. With this degree of freedom, it is straightforward to obtain the 
desired temporal evolution of the metric \cite{ewh}. Therefore, in what
follows, we shall 
analyze the stationary part of Eq. (\ref{traza}).
After separating variables, the radial part of equation (\ref{traza}) can 
be rewritten in the compact form,
\beq
\frac{2 e^{2\Lambda}}{r^2} + \Lambda' \left( \frac{4}{r} +
\frac{\alpha}{r^2}\right) = \left( \frac{2}{r^2} + \frac{\alpha^2}{r^4}
\right).
\label{radial}
\eeq
The change of variables $\Lambda = {\rm ln} \chi$ transforms Eq. 
(\ref{radial}) into a Bernoulli equation, which integrates straightforwardly 
to,
\beq
e^{-2 \Lambda} =\frac{c_{_2} \zeta^2 + c_{_3} \zeta^3 + c_{_4} \zeta^4 +
c_{_5} \zeta^5 + c_{_6} \zeta^6 + c_{_7} \zeta^7 + c_{_8}
\zeta^8 + c_{_9} \zeta^9+  e^{2/\zeta} \zeta^8 {\cal K}\,}
{(4 \zeta +1 )^9} 
\label{lambda}
\eeq
with $\zeta = r/\alpha$,  ${\cal K}$ an integration constant, and the
coefficients $c_{_i}$ are listed in Table I.  
Since the mouths of the wormhole, for a fixed value of $t$, must connect 
two asymptotically flat
spacetimes the
geometry at the wormhole's throat is severely constrained. The so-called 
``flaring out'' condition asserts that the
inverse of the embedding function $r(z)$, must satisfy $d^2 r/dz^2 >
0$ at or near the throat \cite{motho}. Stated 
mathematically, 
\beq
-\frac{\Lambda'\, e^{-2\Lambda}}{(1-e^{-2\Lambda})^2} > 0.
\label{fo1}
\eeq
Moreover, the precise definition of the wormhole's throat (wormhole's 
minimun radius) entails 
a vertical slope of the embedding surface, 
\beq
\lim_{r \rightarrow r\th^+}
\frac{dz}{dr} \, = \lim_{r \rightarrow r\th^+}\, \pm\,
\sqrt{e^{2\Lambda} - 1} = \, \infty.
\label{fo2}
\eeq
Equations (\ref{fo1}) and (\ref{fo2}) will be satisfied if and only if
\beq
\lim_{r \rightarrow r\th^{^+}} e^{-2\Lambda} = 0^+, 
\label{fo}
\eeq
thus, in order to 
fix the constant ${\cal K}$, we must select a value for the 
dimensionless radius ($\zeta\th > 0$) such that Eq. (\ref{fo}) is satisfied. 
As an example, let us impose $\zeta\th$ = 1, 
which sets ${\cal K}  = -\,798517.5 \,e^{-2}$ and
$r\th = \alpha$. It is easily seen that,
as $\zeta \rightarrow \infty $, $e^{2 \Lambda^{^\pm}} \rightarrow 1$.
In this way the nonmonotonic coordinate $\zeta$ (dimensionless
radius) decreases from $+\infty$ to a minimum value $\zeta\th$,
representing the location of the throat of the wormhole and then
increases from $\zeta\th$ to $+\infty$. Because of the ill behavior of 
the radial coordinate 
$r$ near the throat, the spatial geometry is better studied by introducing
the proper radial coordinate, $dl\,=\,
\pm \,e^{\Lambda}\, dr$, which is well behaved throughout the spacetime 
ranging from $-\infty$ to $+\infty$ with $l=0$ at the throat. In this way, if  
$l \rightarrow \pm \,\infty$, $dz/dr \rightarrow 0$, i.e. far from the 
throat in both radial directions the space become asymptotically flat, 
while if $l \rightarrow 0$, 
$dz/dr \rightarrow \infty $, yielding a vertical slope of the 
embedding surface. The aforementioned properties of $\Lambda$, together with 
the definition of $\Phi$, bear out that 
the metric tensor describes two asymptotically flat spacetimes 
joined by a throat for each fixed value of $t$. 
                                                
Concerning the status of WEC's violations, extensions of 
Hawking-Penrose singularity 
theorems for EC theory have been already performed \cite{k}. The inequality,  
\beq
\left(\, \tilde{\sigma}^{\mu\nu} - {1 \over 2} \, g^{\mu\nu}\, 
\tilde{\sigma}_\eta\,^\eta\,
\right)\,
\xi_\mu\, \xi_\nu \geq 0 
\eeq 
that must hold for all timelike unit vectors 
$\xi^\mu$, generalizes
the strong energy condition in EC theory. The corresponding
generalization for WEC is given by the inequality,
\beq
\tilde{\sigma}^{\mu\nu} \xi_\mu\, \xi_\nu \geq 0 
\label{WEC}
\eeq
that must hold for all timelike unit vectors 
$\xi^\mu$.  
It is easily
seen following the procedure sketched in \cite{ewh} that the
dynamics of the geometry lets one move the energy condition violating
region around in time leading to a temporary suspension of the need for the 
WEC violation. This hardly solves the problem since struts of ``exotic'' 
material are still required at least in an infinitesimal time interval.   
It is possible to address this problem in $U_4$ theory with a 
distribution of matter with square spin effects overwhelming 
the mass terms. Hehl et al \cite{hel} have 
estimated the required critical 
density $\rho_{\rm_C}$ for such a situation. 
In the case of a spin fluid of neutrons with isotropic pressure, 
$\rho_{\rm _C} \approx 10^{54}$ g cm$^{-3}$, 
which is orders of magnitude larger than the
density at the center of the most massive neutron stars. 
In the case of electrons, the estimated value is
$\rho_{\rm _C} \approx 10^{47}$ g cm$^{-3}$.

\section{Outlook}

It was shown that EC theory admits evolving wormhole geometries
which could hide a ``flash'' of exoticity in the torsion tensor. 
In spite of the 
very restrictive conditions upon the matter density, the very early universe 
might provide a fruitful scenario for wormholes to rise.  Unfortunately, 
the functional forms of the conformal factor $\Omega (t)$ 
associated with realistic descriptions 
of the expansion of the universe cannot be used for the enlargement of 
these wormholes (see the discussion about spherically symmetric 
time-dependent wormholes of ref. \cite{haifa}).

\acknowledgements

This work has been partially suported by FOMEC.

\begin{table}
\caption{Coefficients of the function $e^{-2 \Lambda}$}
\begin{tabular}{ccccccccc}
$c_{_2}$  & $c_{_3}$ & $c_{_4}$ & $c_{_5}$ & $c_{_6}$ & $c_{_7}$ 
& $c_{_8}$ & $c_{_9}$ \\ \hline
2 & 70 & 1071 & 9310 & 49805 & 164493 & 311622.5 & 262144 
\end{tabular}
\end{table}

\end{document}